\documentclass[conference]{IEEEtran}
\IEEEoverridecommandlockouts
\usepackage{cite}
\usepackage{amsmath,amssymb,amsfonts}
\usepackage{algorithmic}
\usepackage{graphicx}
\usepackage{textcomp}
\usepackage{xcolor}
\usepackage{booktabs}
\usepackage{subcaption}
\usepackage{comment}
\usepackage{tcolorbox}
\usepackage[bookmarks=false]{hyperref}
\usepackage{soul}
\newtcolorbox{todo}[2][]{colbacktitle=red!10!white,
	colback=blue!10!white,coltitle=red!70!black,
	title={#2},fonttitle=\bfseries,#1}

\makeatletter
\newcommand{\linebreakand}{%
\end{@IEEEauthorhalign}
\hfill\mbox{}\par
\mbox{}\begin{@IEEEauthorhalign}
}

\def\BibTeX{{\rm B\kern-.05em{\sc i\kern-.025em b}\kern-.08em
    T\kern-.1667em\lower.7ex\hbox{E}\kern-.125emX}}
\begin{document}


\newif\ifcomments %
\commentstrue 

\newcommand{\prasad}[1]{%
  \ifcomments%
    \textit{\textbf{\small{\textcolor{red}{[PRASAD: #1]}}}}
  \else%
  \fi%
}

\newcommand{\pgl}[1]{%
  \ifcomments%
    \textit{\textbf{\small{\textcolor{blue}{[PGL: #1]}}}}
  \else%
  \fi%
}

\newcommand{\phm}[1]{%
  \ifcomments%
    \textit{\textbf{\small{\textcolor{purple}{[PHM: #1]}}}}
  \else%
  \fi%
}

\newcommand{\sgil}[1]{%
  \ifcomments%
    \textit{\textbf{\small{\textcolor{blue}{[SGIL: #1]}}}}
  \else%
  \fi%
}

\newcommand{\ek}[1]{%
  \ifcomments%
    \textit{\textbf{\small{\textcolor{brown}{[Eduard: #1]}}}}
  \else%
  \fi%
}

\newcommand{\cg}[1]{%
  \ifcomments%
    \textit{\textbf{\small{\textcolor{purple}{[CG: #1]}}}}
  \else%
  \fi%
}

\title{Digital Twin as a Service (DTaaS):\\ A Platform for Digital Twin Developers and Users}

\author{\IEEEauthorblockN{Prasad Talasila\textsuperscript{10000-0002-8973-2640}}
	\IEEEauthorblockA{prasad.talasila@ece.au.dk}
	\and
	\IEEEauthorblockN{Cláudio Gomes\textsuperscript{10000-0003-2692-9742}}
	\IEEEauthorblockA{claudio.gomes@ece.au.dk}
	\and
	\IEEEauthorblockN{Peter Høgh Mikkelsen\textsuperscript{10000-0003-2321-758X}}
	\IEEEauthorblockA{phm@ece.au.dk}
	\linebreakand
	\IEEEauthorblockN{\hspace*{0.07in}}
	\and
	\IEEEauthorblockN{Santiago Gil Arboleda\textsuperscript{10000-0002-1789-531X}}
	\IEEEauthorblockA{sgil@ece.au.dk}
	\and
	\IEEEauthorblockN{Eduard Kamburjan\textsuperscript{20000-0002-0996-2543}}
	\IEEEauthorblockA{eduard@ifi.uio.no}
	\and
	\IEEEauthorblockN{Peter Gorm Larsen\textsuperscript{10000-0002-4589-1500}}
	\IEEEauthorblockA{pgl@ece.au.dk}
	\linebreakand
	\IEEEauthorblockN{\hspace*{0.1in}}
	\and
	\IEEEauthorblockA{\textsuperscript{1}
		\textit{Centre for Digital Twins, DIGIT} \\
		\textit{Department of ECE} \\
		\textit{Aarhus University} \\
		\textit{Finlandsgade 22, 8200 Aarhus N, Denmark}}
	\and
	\IEEEauthorblockA{\textsuperscript{2}
		\textit{Department of Informatics, University of Oslo} \\
		\textit{Gaustadalleen 23 B, 0373 Oslo, Norway}}
}

\maketitle

\begin{abstract}
Establishing digital twins is a non-trivial endeavour especially when users face significant challenges in creating them from scratch. Ready availability of reusable models, data and tool assets, can help with creation and use of digital twins. A number of digital twin frameworks exist to facilitate creation and use of digital twins. In this paper we propose a digital twin framework to author digital twin assets, create digital twins from reusable assets and make the digital twins available as a service to other users. The proposed framework automates the management of reusable assets, storage, provision of compute infrastructure, communication and monitoring tasks. The users operate at the level of digital twins and delegate the rest of the work to the digital twin as a service framework.
\end{abstract}


\begin{NoHyper}
	
\section{Introduction}

Digital Twins (DTs) are used to add value to systems of interest which typically are called Physical Twins (PTs). 
Such DTs can be assisting individual Cyber-Physical Systems (CPSs) in different ways and it is the vision of these capabilities that makes DTs so interesting. At the heart of a DT is a collection of models describing characteristics of the CPS of interest. In practice, a comprehensive DT may marshal a collection of diverse models, each developed for different purposes. These models are evaluated on tools using the data received from PT.  Delivering such a DT can be a complex task, especially because of the need for several different modelling approaches, including information models, geometry, physics, and behaviours~\cite{Tao2022, Zambrano22}. 

The main motivation for DTs is to enable real-time monitoring, analysis, and simulation of a PT. This technology facilitates improved decision-making, predictive maintenance, and optimization in various industries, including manufacturing, healthcare, and urban planning. DTs enhance efficiency, reliability, and sustainability by providing a comprehensive understanding of complex systems and supporting data-driven insights. They represent a natural stepping stone from the massive availability of sensors and data in different industries, and it is our conviction that their architecture is common across many such industries.

A number of DT platforms have been proposed to reduce the implementation effort in relation to the structural aspects of DTs~\cite{Lehner2021a}.
Nevertheless, the other functional and behavioural aspects are not necessarily supported. 
Other approaches for co-simulation-oriented DTs can be complementary to existing DT platforms~\cite{Fitzgerald2019}, although they may need further contextual design for DT structures to achieve reusability. 

Recently the idea of establishing a Digital Twin as a Service (DTaaS) has been proposed but with a primary focus on services related to Augmented Reality~\cite{Aheleroff21}. In this paper we extend that vision with more detail about the DT assets involved and their realisation in a more general context. We describe the foundational concepts needed to create DT platforms. These concepts are then used to define a candidate system architecture for a DTaaS platform. The primary system requirement for the proposed DT platform is to support DT through different phases of its lifecycle. The create, execute, save, analyse, evolve and terminate phases of DT lifecycle are supported. One possible implementation of the proposed system is described in this paper. A library of reusable assets reduces the difficulties faced by users in creation of DTs for different CPSs~\cite{Human2023}. The DTaaS platform supports DTs that can be created from readily available and reusable library assets. Such a reuse facilitates adoption of DTs among even the non-technical users.

The rest of this paper is structured such that Section~\ref{sec:background} presents the background necessary to understand the results reported here. Afterwards, Section~\ref{sec:dtlife} presents the envisaged lifecycle of using DTs. This is followed by an overview of the system architecture in Section~\ref{sec:architecture}. Then an overview of the corresponding implementation is presented in Section~\ref{sec:impl}. Finally, Section~\ref{sec:future} and Section~\ref{sec:conclude} completes the paper with future work and concluding remarks respectively.
\section{Background}\label{sec:background}

A DT is digital representation of a physical object, that through data exchange, reflects evolution of a PT over time \cite{Grieves17}.
Kritzinger et al. \cite{Kritzinger2018} propose three subcategories of DTs: (i) Digital Model, in which PT and DT have no automated exchange of data, (ii) Digital Shadow, where PT emits data to the DT automatically and finally (iii) Digital Twin, in which case there exist a two-way automated data exchange between the DT and PT. 

The composition of a DT is considered by \cite{Lee2015} who proposes a five-level architecture. On the lowest level (i) Smart Connection, resides the data exchange between DT and PT; (ii) Data-to-information conversion, concerns conversion and aggregation of data for monitoring and to make it useful for (iii) Cyber, which is the central information hub and source of analysis across multiple data sources. At level (iv) Cognition, the knowledge acquired from lower levels are made available for decision-making and finally at level (v) Configuration, is where decisions or reconfiguration from the DT is fed back to the PT, to make it self-adaptable. Thus, to create a DT, infrastructure, tools, models and configurations must address each level.

DTs might often be engineered and operated by reusing existing software and parts of the DT.
Completely new development of a DT for each PT is not necessary if the DT or its parts are reusable \cite{Dalibor2022}. A DT can be structured to be composed of reusable assets one of which is a model. Zambrano et. al \cite{Zambrano22} discuss the reusability of models. The DIGITbrain platform focuses on reusable data, models, and algorithms (alias for software tools and frameworks). The platform considers four reusable assets for creating DTs: data, models, algorithms (tools) and model-algorithm pairs \cite{Talasila2022}.  

The reusable assets are to be selected and configured to create new DTs. The configuration of DTs is also addressed in \cite{Dalibor2022} which uses Domain Specific Languages (DSLs) to describe domain, data, tagging, constraint, and GUI as input to a set of generation tools, to create the application-independent parts of a DT platform. The platform can then be tailored by a domain expert for a specific purpose, using a web-based frontend. Several commercial and open-source projects propose DSLs to specify composition of a DT as a configuration of models, relations and data exchange \cite{Pfeiffer2022}.


Existing development frameworks for DTs use different modelling approaches, which require different sets of tools and functional aspects as long as the bidirectional connectivity is complied. They are normally based on reference architectures for the definition of schemas, functionalities, and services. 
These frameworks usually extend the infrastructure from Internet of Things (IoT) frameworks \cite{Bader2020}.

The first group is based on object-oriented design \cite{Lehner2021a}. 
The second group is based on co-simulation as a backbone of the DTs. In this group, the frameworks use behavioural models embedded in Functional Mock-up Units (FMUs). The interfaces to inputs and outputs need to be provided externally or by hand-coding.
An example of this group is the INTO-CPS Co-simulation Framework \cite{Thule2019}, which is composed of an orchestration engine for FMUs and provide some additional FMUs for connectivity.

Others, such as DIGITbrain and HUBCAP have used a model-based design (MBD) approach in order to enable the reusability of DT assets from a high-level perspective \cite{Talasila2022}. These assets can be extended to a DTaaS platform by considering replaceable tools and other aspects, such as interaction with PTs,  users, and semantic data exchange~\cite{Aheleroff21}.

The DTaaS platform is a complementary approach across different DT frameworks.
It provides the infrastructure to run and maintain DTs with a well-defined DT asset configuration. The existing DT frameworks can be run within the DTaaS software including the integration of live DTs running within the software to external services. It means that, other existing DT frameworks can make use of the reusable components and services that the DTaaS software offers.
On the other hand, the DTaaS platform offers two additional features that are not considered in existing DT frameworks that can be highly beneficial for users, which are related to i) providing private workspaces for authoring and verification of reusable assets and ii) extending from the private workspaces, users are also allowed to collaborate and share models and assets in the platform, which enables the reusability of developed assets by other users.

A key concern in all the frameworks is to support different operating phases of DT. Successful software engineering practices such as microservices, DevOps and GitOps have found their way into the development and operation of DT platforms \cite{app11125633, Ciavotta2019}. The DT platforms hosting reusable DT assets must be able to verify the promised functionality of the assets. The DevOps practices such as continuous integration can help with the verification of published assets \cite{reiterer2020continuous}. The GitOps practices \cite{beetz2021gitops} allow provisioning of DT execution infrastructure using familiar git workflows. These practices can be adopted to support scalable execution of DTs.

In the rest of the paper, the software platform and the software phrases are used to refer to the DTaaS software. 


\begin{figure*}
	\centering
	\begin{subfigure}{0.32\textwidth}
		\includegraphics[width=1.0\textwidth]{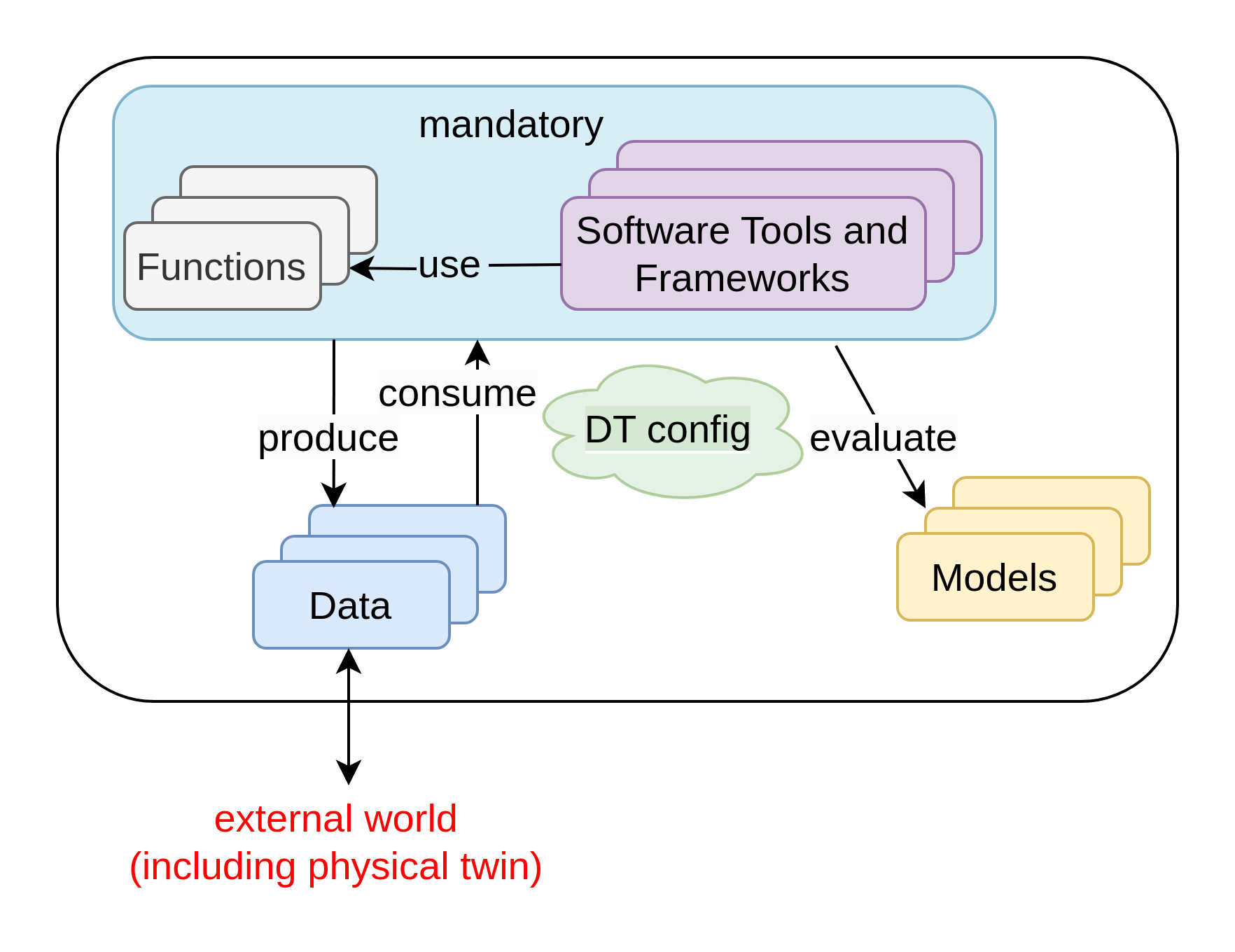}
		\caption{Constitution of a minimum viable DT. At least one asset from mandatory set is required. An executable binary asset is a valid DT.}
		\label{fig:assets}
	\end{subfigure}
	\hspace*{12pt}
	\begin{subfigure}{0.32\textwidth}
		\includegraphics[width=1.0\textwidth]{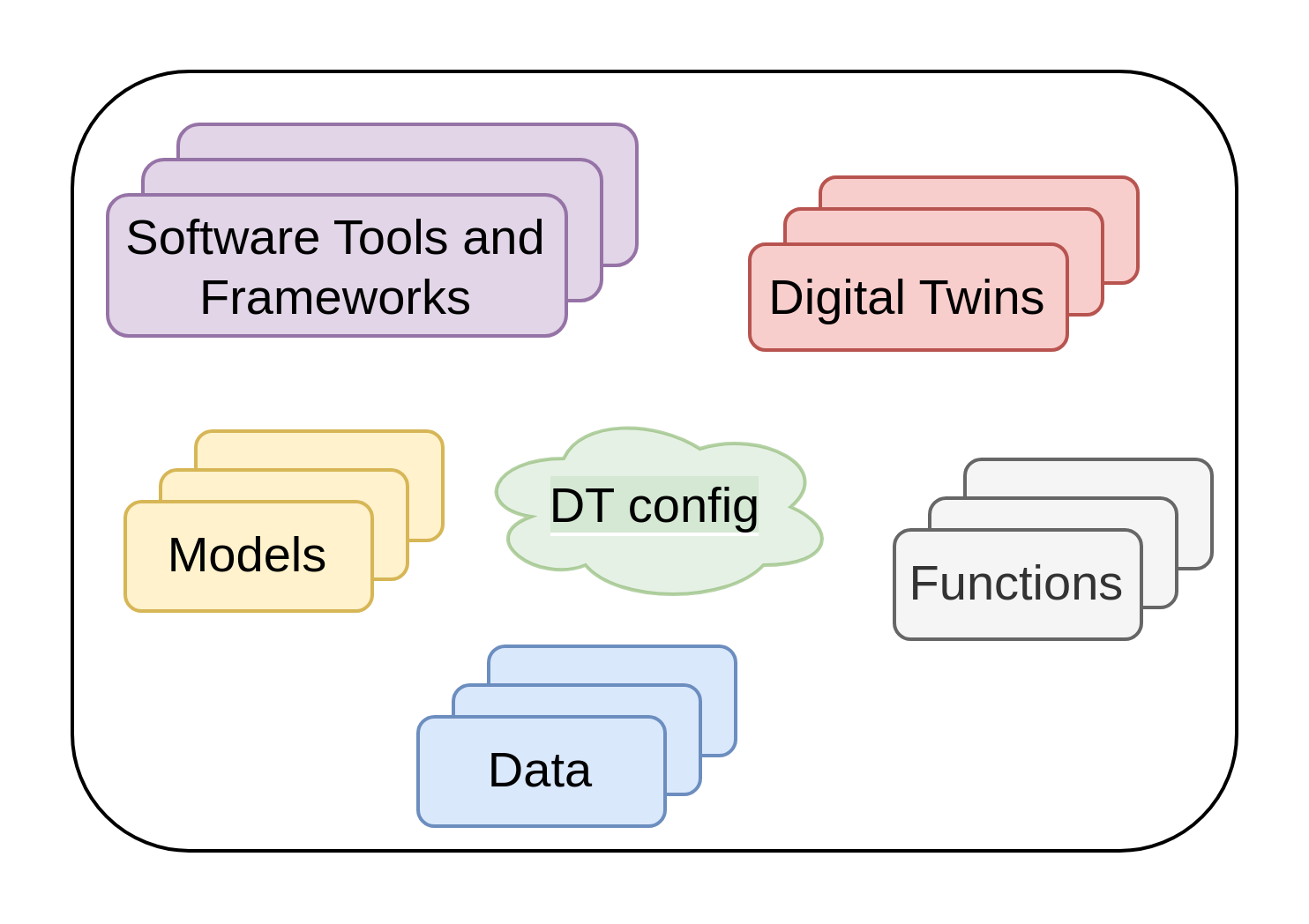}
		\caption{Composability among DTs. Composition introduces recursion in the construction of DTs.}
		\label{fig:composability}
	\end{subfigure}
	\hspace*{12pt}
	\begin{subfigure}{0.25\textwidth}
		\includegraphics[width=1.1\textwidth]{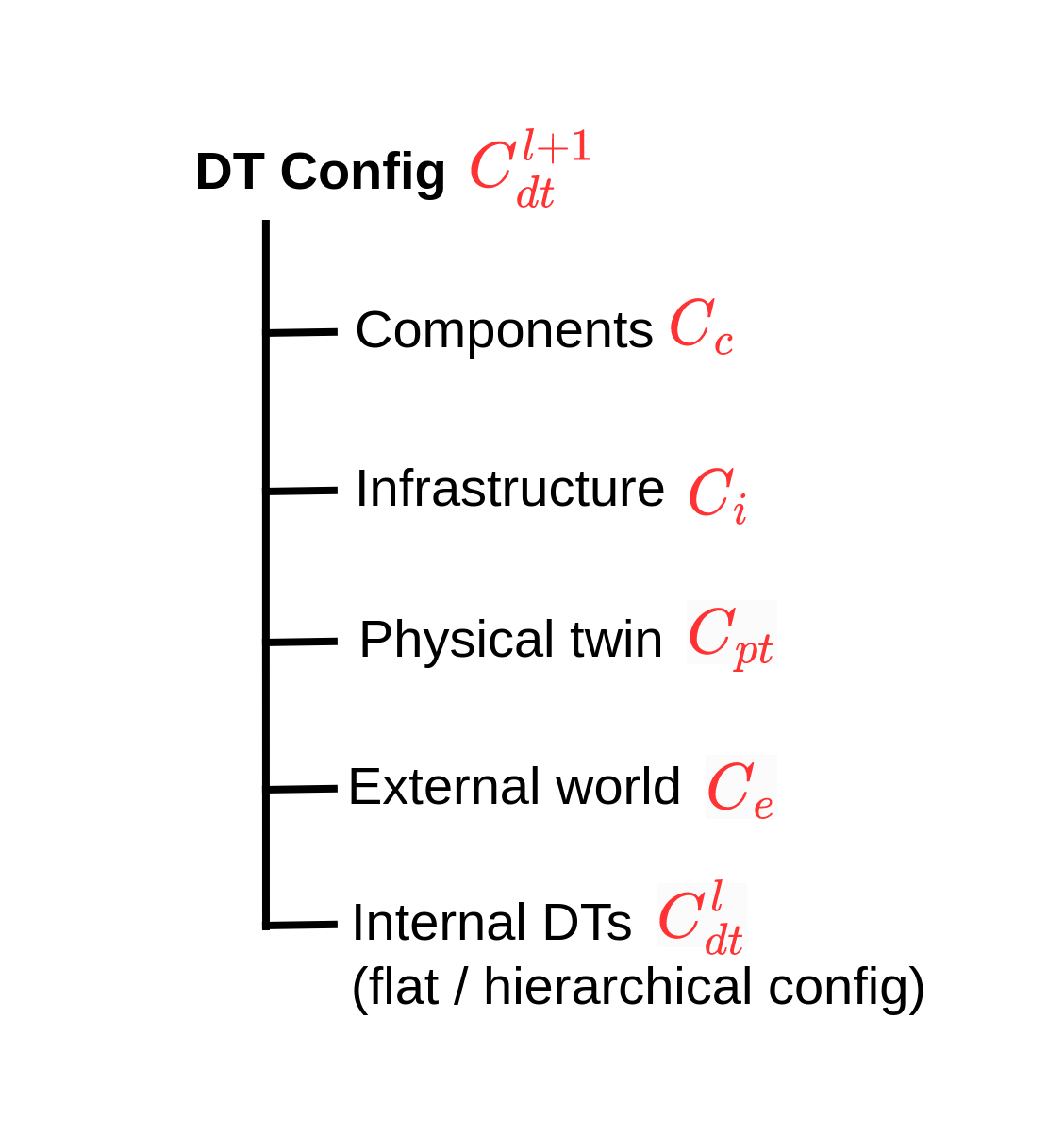}
		\caption{Configuration possibilities for DTs. Flat configuration is more flexible yet more complex.}
		\label{fig:config}
	\end{subfigure}
	
	\caption{Data, Model, Function, Tool, ready to use Digital Twin assets of the DT platform. The DT configuration links assets to form a meaningful DT.}
	\label{fig:dt-properties}
\end{figure*}

\section{The Digital Twin Lifecycle} \label{sec:dtlife}

\subsection{DT Assets}\label{sec:dt-assets}

The DTaaS software platform treats DTs as having reusable assets. These assets are put together and configured in a certain way. We use four categories of assets: data (D), model (M), function (F) and Tool (T).

The data (D) asset refers to data sources and sinks available to a DT. Typical examples of data sources are sensor measurements from the PT, and test data provided by manufacturers for calibration of models. Typical examples of data sinks are visualisation software, external users and data storage services. There exist special outputs such as events, and commands which are akin to control outputs from a DT. These control outputs usually go to the PT, but they can also go to another DT~\cite{Esterle2021}.

The model (M) assets are used to describe different aspects of a PT and its environment, at different levels of abstraction. Therefore, it is possible to have multiple models for the same PT. For example, a flexible robot used in a car production plant may have structural model(s) which will be useful in tracking the wear and tear of parts. The same robot can have a behavioural model(s) describing the safety guarantees provided by the robot manufacturer. The same robot can also have a functional model(s) describing the part manufacturing capabilities of the robot.

The function (F) assets are primarily responsible for pre- and post-processing of: data inputs, data outputs, control outputs. The research results from data science can be used to create useful function assets for the platform.
In some cases, DT models require calibration prior to their use; functions written by domain experts along with right data inputs can make model calibration an achievable goal. Another use of functions is to process the sensor and actuator data of both the PT and the DT.

The software tool (T) assets are software used to create, evaluate and analyse models. These tools are executed on top of a computing platforms, i.e., an operating system, or virtual machines like Java virtual machine, or inside docker containers. The tools tend to be platform specific, making them less reusable than models. 
A tool can be packaged to run on a local or distributed virtual machine environments thus allowing selection of most suitable execution environment for a DT. 
Most models require tools to evaluate them in the context of data inputs. 
There exist cases where executable packages are run as binaries in a computing environment. Each of these packages are a pre-packaged combination of models and tools put together to create a ready to use DT.

\begin{figure*}[ht]
	\centering
	\includegraphics[width=0.9\textwidth]{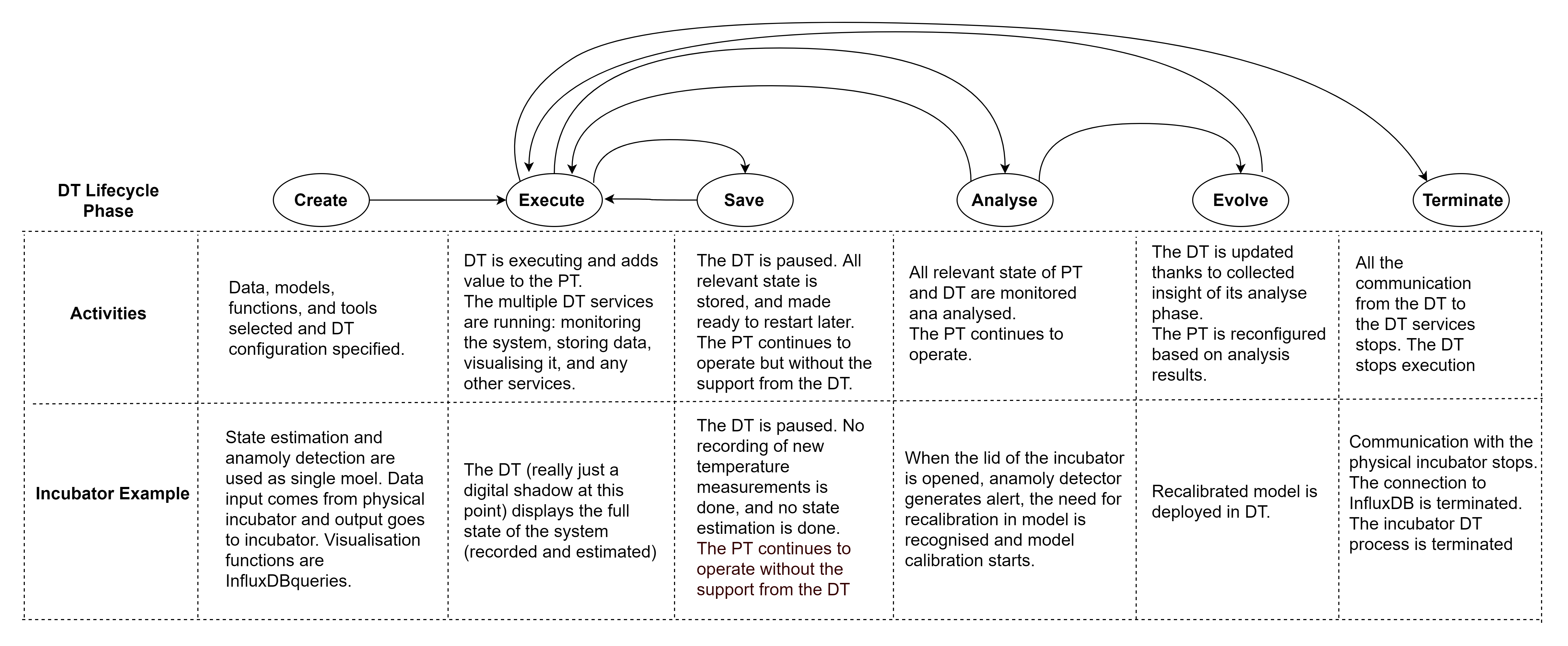}
	\caption{Mapping of the DT lifecycle phases to the incubator use case \cite{Feng2021}.}
	\label{fig:lifecycle-incubator-mapping}	
\end{figure*}

There is a dependency between the assets especially in the context of creating DTs. These dependencies are illustrated in Figure~\ref{fig:assets}. 
Only functions/tools can use models/data. 
A specific combination of these assets constitute a DT. The interconnections between assets, parameters (configurable run-time variables) of the assets need to be specified for each DT. This information becomes a part of the DT asset configuration ($C_{a}$). The information encoded in $C_a$ is not sufficient to truly manifest a closed loop communication between a DT and a PT. Thus each DT requires complete configuration ($C_{dt}$) that is sufficient to execute a DT in the presence of supporting services and execution environment. The possibilities of asset combinations used in a DT can be expressed using Equation~\ref{eq:dt-def}. 
\begin{align}\label{eq:dt-def}
	D_t: & \{ D{^*},M^{*},(FT)^{+} \}C_{dt}
\end{align}
\noindent where D denotes data, M denotes models, F denotes functions, T denotes tools, $C_{dt}$ denotes DT configuration and $D_t$ is a symbolic notation for a DT itself. The  $\{ D{^*},M^{*},(FT)^{+} \}C_{dt}$ expression denotes composition of DT from D,M,T and F assets. The $*$ indicates zero or one more instances of an asset and $+$ indicates one or more instances of an asset.

\subsection{DT Configuration}\label{sec:dt-config}
The DT asset configuration ($C_{a}$ introduced in Section~\ref{sec:dt-assets}) is only a part of the complete configuration ($C_{dt}$) needed for running a live DT with feedback loop to its PT. The $C_{pt}$ denotes configuration information required by a DT to communicate with a PT. Each DT may have constraints on the kind of execution environments it is capable of using, i.e. tools that can only run either on a specific operating system or on a server with specific hardware capabilities. The $C_i$ denotes the infrastructure configuration required by a DT. The $C_e$ denotes configuration for integration of a DT with external software systems, ex. third-party visual dashboards.
\begin{equation}\label{eq:dt-config}
	C_{dt} = \{ C_a, C_i, C_e, C_{pt}\}
\end{equation}
Among all the configurations shown in Equation~\ref{eq:dt-config}, $C_a$ and $C_{pt}$ are very specific to one DT or a class of DTs. Thus, generalization of these two configurations into a configuration specification standard is a challenging task. The other two configurations -- $C_i$ and $C_e$ -- are more general and a configuration specification standard for these two is a manageable challenge. A sanity check is required on validity of any given $C_{dt}$.


Two situations demand adjustments to $C_{dt}$. One is a user-driven change in $C_{a}$, $C_{pt}$, $C_i$, $C_e$, or $C_{dt}$ of included DTs. In this case, a validity check is required before a transition to new a configuration can be made. Second is a requirement to perform a what-if analysis. 
A what-if analysis requires minor variations on $C_{dt}$ to plan and optimize future steps to be undertaken either on a PT or a DT. 
Actual implementation of a what-if analysis can be resource intensive with the resource requirements scaling up in proportion to algorithmic bounds on the (sub)-systems being used by a DT.

A DT can also use external tools such as planning and optimization. This is especially true in what-if analysis. If these tools are used exclusively within a DT, then they can be considered as tools in asset library. Otherwise, they are part of the infrastructure / external world.

\subsection{Phases in DT Lifecycle}
\label{sec:dt_phases}

A DT lifecycle consists of \textit{create, execute, save, analyse, evolve} and \textit{terminate} phases. In addition to having D,M,T and F as assets in the library, the ready to use DTs can also be a library assets. 
Users might choose to create a new DT or select an existing DT. The \textit{create} phase involves asset selection and an specifying DT configuration.   Sections~\ref{sec:dt-assets} and~\ref{sec:dt-config} describe the configuration phase. If DT is reused, there is no creation phase at the time of reuse. 

The \textit{execute} phase involves automated execution of a DT based on its configuration. 
If a DT is reused, there will be a temporal gap between creation and execution times of a DT. Thus, a need might arise for just-in-time DT reconfiguration at the point of execution. The \textit{save} phase involves saving the state of DT to enable future recovery. The \textit{terminate} phase involves stopping the execution of DT and releasing all the resources and connections mentioned in the DT configuration.

\paragraph*{Analyse and Evolve Phases}
Monitoring, at its most basic level, requires data gathering and storage, of the interaction between the PT and its environment, and among the PT's assets.
However, even for a simple system such as an incubator (like the one introduced in \cite{Feng2021}), monitoring requires that hidden quantities (that is, quantities for which we cannot obtain a sensor measurement directly) be estimated.
This activity corresponds to the \textit{analyse} lifecycle phase.
Often these hidden quantities are represented by variables in the various models used by the DT.
The consequence is that estimates of these hidden quantities need to be stored in the database, becoming then input to decision-making simulations, where all variables of the models need to be properly initialized.  For more details we refer the reader to \cite{Feng2021a}.


\begin{figure*}
	\centering
	\includegraphics[width=0.8\textwidth]{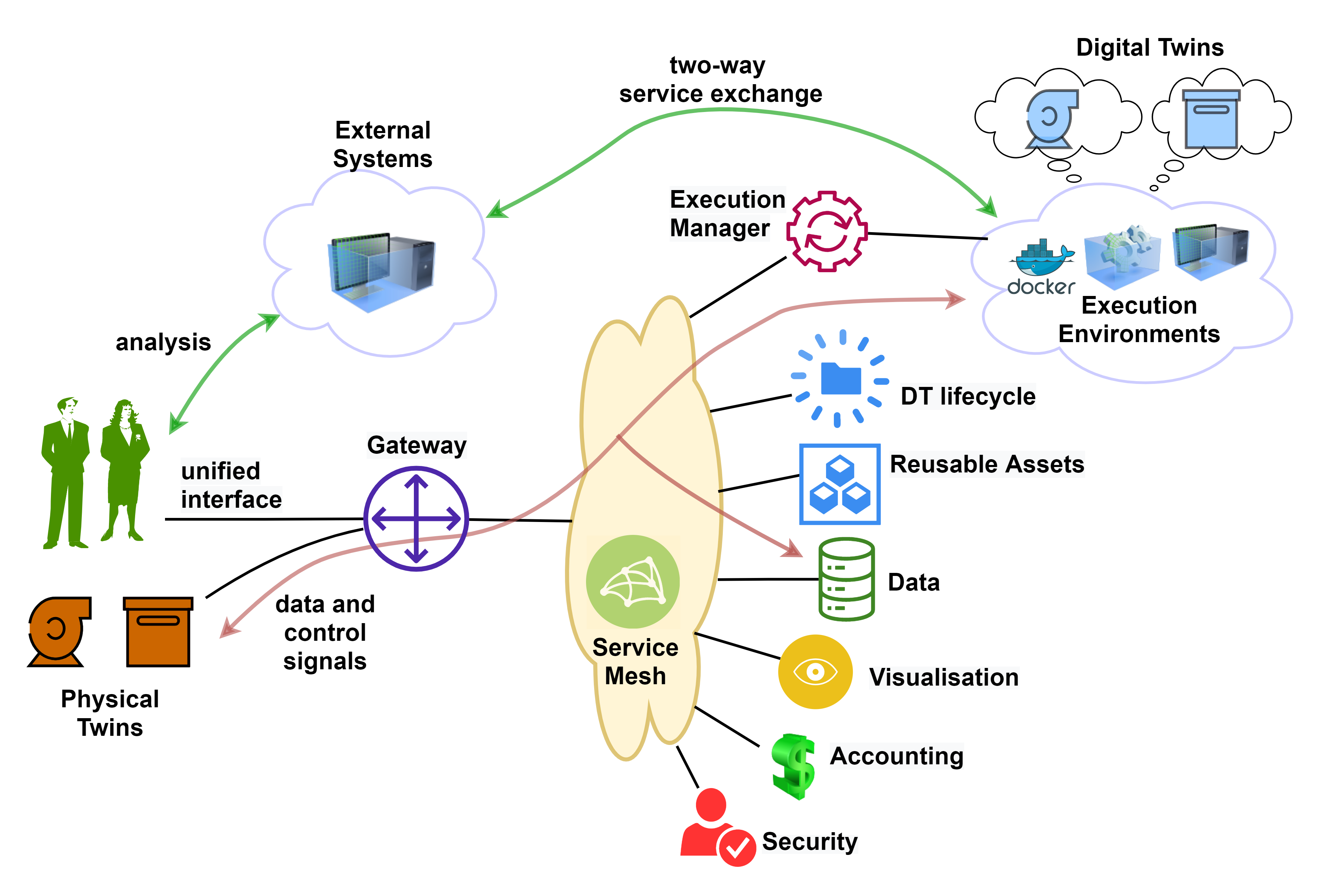}
	\caption{\centering System components required to fulfil the functional requirements of a DTaaS software.}
	\label{fig:architecture}
\end{figure*}

Monitoring also informs the next lifecycle stage of the DT: the \textit{evolve} stage. The evolve phase involves user/event-triggered reconfiguration of an instantiated DT. 
Note that the monitoring and planning steps make use of the other DT services.


\paragraph*{Reconfiguration Phase and Consistency}
The evolution phase requires reconfiguration of DTs. The aim of reconfiguration is to ensure \emph{consistency} between DT and its PT,  i.e., the adequacy of the DT to mirror its PT, access its data, and enable the required analyses.

Reconfigurations may be triggered by different kinds of events, two of which we discussed above, they are specific to the system and, thus, reconfiguration procedures must be provided by the user. These procedures are highly application specific. The reconfiguration procedures must be able to access the current DT configuration and its assets. As the configuration is highly heterogeneous, the platform should offer a way for \emph{uniform} access to it, i.e., a representation mapping $\mu$ that is defined on $C_{dt}$ \emph{and all its assets} as well as provide an interface for the user to program reconfigurations in terms of the uniform access, i.e., define transitions
$\mu\left(C^{l}_{dt}\right) \rightarrow_\mu \mu\left(C^{l+1}_{dt}\right).$
Knowledge graphs are a suitable technology to implement for the mappings. The approaches to express consistency between DT and PT in terms of queries on knowledge graphs have been shown to be useful~\cite{KamburjanJ22}.

\begin{figure*}
	\centerline{\includegraphics[scale=0.1]{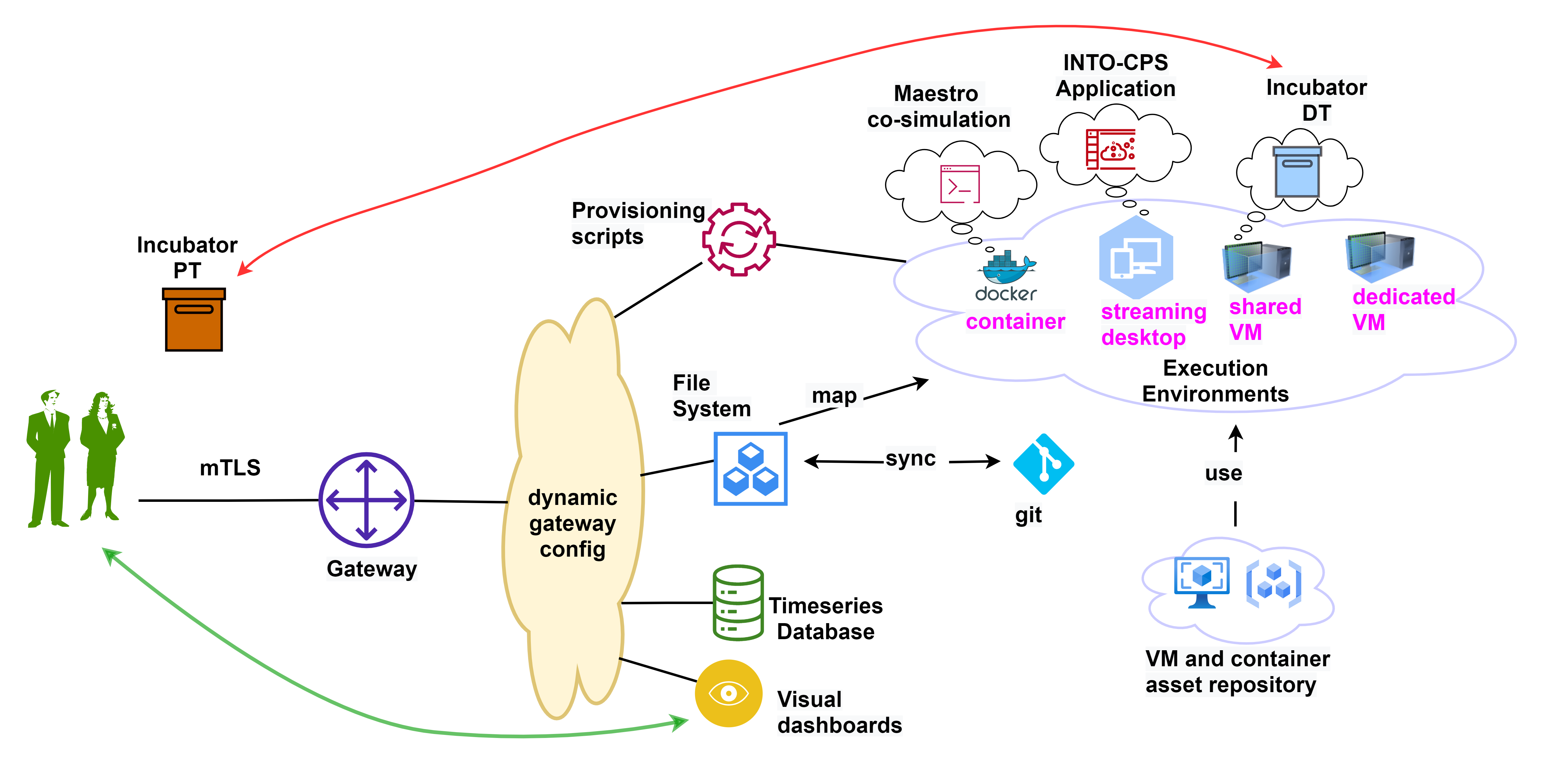}}
	\caption{Current implementation catering to co-simulation and monolithic DTs. The system supports creation of DTs from library assets.}
	\label{fig:current-status}
\end{figure*}


\subsection{The Incubator Use Case}
Here we introduce a simple incubator as an example to make the description of DT lifecycle stages more clear. Consider an incubator that we wish to make smart by the use of a DT.
The PT consists of an insulated box with a heater inside, that can be turned on or off by a controller, which measures the temperature inside. Here the DT can adjust the parameters of the controller based on past performance data, and detect/mitigate anomalies, such as the box being opened unexpectedly. Feng et. al \cite{Feng2021a,Feng2022a} provide complete details on this use case.
Figure~\ref{fig:lifecycle-incubator-mapping} contains a mapping of different activities done for the  incubator DT and the conceptual description of DT lifecycle phases.

\subsection{Hierarchical DTs}
\label{sec:hierarchical-dt}

A PT like a manufacturing factory consists of multiple robots each of which can be have a DT of their own. This hierarchical aggregation is to replicate the compositional properties of a PT. A hierarchical DT can also use the elementary DT assets in addition to the aggregated DTs \cite{Esterle2021}. Figure~\ref{fig:composability} shows this possibility. Thus Equation~\ref{eq:dt-def} gets modified to become:
\begin{align}\label{eq:hierarchical-dt}
	D_t: & \{ D{^*},M^{*},(FT)^{+} \}C_{dt}\\
	& \{ (DMFT)^{*},D^{+}_{t} \}C_{dt}	\notag
\end{align}
\noindent where the $\{ (\mathit{DMFT})^{*},D^{+}_{t} \}C_{dt}$ expression represents use of one DT and other reusable assets to create another DT.

In the case of hierarchical DTs, there may be a need to provide configuration ($C_{dt}$) for each of the DTs included. Thus, Equation~~\ref{eq:dt-config}  becomes generalized to Equation~\ref{eq:hierarchical-dt-config} for a hierarchical DT configuration. In case of non-hierarchical DTs, $l = 0$ and $C^{l}_{dt} = \phi$. 
\begin{equation}\label{eq:hierarchical-dt-config}
	C^{l+1}_{dt} = \{ C_a, C_i, C_e, C_{pt},C^{l}_{dt} \}
\end{equation}

Figure~\ref{fig:config} shows a graphical illustration of the relationship between $C_{dt}$ and all other configurations.

The DT lifecycle of composed DT ($D^{l}_{t}$) is dictated by the composing DT ($D^{l+1}_{t}$). When $D^{l+1}_{t}$ goes through Execute, Persist and Terminate phases the $D^{l}_{t}$ must adhere to the lifecycle phase of $D^{l+1}_{t}$. Otherwise, the lifecycle phases of $D^{l+1}_{t}$ and $D^{l}_{t}$ are independent.

\section{System Architecture}\label{sec:architecture}

\subsection{Requirements}

The DTaaS software platform users expect a single platform to support the complete DT lifecycle. To be more precise, the platform users expect the following features:
\begin{enumerate}
	\item \textbf{Author} -- create different assets of the DT on the platform itself. This step requires use of some software frameworks and tools whose sole purpose is to author DT assets.
	\item \textbf{Consolidate} -- consolidate the list of available DT assets and authoring tools so that user can navigate the library of reusable assets. This functionality requires support for discovery of available assets.
	\item \textbf{Configure} -- support selection and configuration of DTs. This functionality also requires support for validation of a given configuration.
	\item \textbf{Execute} -- provision computing infrastructure on demand to support execution of a DT.
	\item \textbf{Explore} -- interact with a DT and explore the results stored both inside and outside the platform. Exploration may lead to analytical insights.
	\item \textbf{Save} -- save the state of a DT that's already in the execution phase. This functionality is required for on-demand saving and re-spawning of DTs.
	\item \textbf{What-if analysis} -- explore alternative scenarios to (i) plan for an optimal next step, (ii) recalibrate new DT assets, (iii) automated creation of new DTs or their assets; these newly created DT assets may be used to perform scientifically valid experiments.
	\item \textbf{Share} -- share a DT with other users of their organisation.
\end{enumerate}

\subsection{System Components}
Despite the different user requirements, the platform must present a unified interface to the users. This unified interface is achieved by providing a gateway to consolidate the functionality provided by internal system components. 
Figure \ref{fig:architecture} shows the system architecture of the the DTaaS software platform. 
The users interact with the software platform using a website. The gateway is a single point of entry for direct access to the platform services. The gateway is responsible for controlling user access to the microservice components. The microservices are complementary and composable; they fulfil core requirements of the system. The service mesh enables discovery of microservices, load balancing and authentication functionalities. There are microservices for catering to author, store, explore, configure, execute and scenario analysis requirements.

\subsection{Microservices}
The microservices illustrated in Figure~\ref{fig:architecture} provide bulk of the platform functionality. The security microservice implements role-based access control (RBAC) in the platform. The accounting microservice is responsible for keeping track of the platform, DT asset and infrastructure usage. Any licensing, usage restrictions need to be enforced by the accounting microservice. Accounting is a pre-requisite to commercialisation of the platform. Due to significant use of external infrastructure and resources via the platform, the accounting microservice needs to interface with accounting systems of the external services.

The data microservice is a frontend to all the databases integrated into the platform. A time-series database and a graph database are essential. These two databases store time-series data from PT, events on PT/DT, commands sent by DT to PT. The PTs uses these databases even when their respective DTs are not in the execute phase.

The visualisation microservice is again a frontend to visualisation software that are natively supported inside the platform. Any visualisation software running either on external systems or on client browsers do not need to interact with this microservice. They can directly use the data provided by the data microservice.

The reusable assets microservice (Asset MS) provides search, explore, and select functions over DT assets. Thus Asset MS should aid users in performing create-read-update-delete operations on the private and shared reusable assets. Any ready to use DTs are also made available via the Asset MS.

The execution manager microservice (Exec MS) is responsible for on-demand provisioning of virtual compute infrastructure. To make the platform scalable, the Exec MS must be capable of integrating with private and public cloud providers. The Exec MS creates virtual workspaces on top of the provisioned compute infrastructure.  Users operate with these isolated workspaces.

The DT lifecycle microservice assists users during all phases of a DT. This microservice extensively uses other microservices to provide atomic operations at the level of DTs. This microservice acts as a controller to both the Asset and Exec microservices.

\section{Implementation}\label{sec:impl}

The DTaaS software platform is currently under development. Crucial system components are in place with ongoing development work focusing on increased automation and feature enhancement. Figure~\ref{fig:current-status} shows the current status of the development work. The current security functionality is based on signed Transport Layer Security (TLS) certificates issued to users. The TLS certificate based mutual TLS (mTLS) authentication protocol provides better security than the usual username and password combination. The mTLS authentication takes place between the users browser and the platform gateway. The gateway federates all the backend services. The service discovery, load balancing, and health checks are carried by the gateway based on a dynamic reconfiguration mechanism.

A time-series database is now incorporated in the platform. The database comes with query and visualisation dashboards. Users are permitted to share the dashboards. Thus DT experts can develop custom dashboards and share them with other users. A file server has been setup to act as a DT asset repository. Each user gets space to store private DT assets and also gets access to shared DT assets. Users can synchronize their private DT assets with external git repositories. In addition, the asset repository transparently gets mapped to user workspaces within which users can perform DT lifecycle operations.

All users have dedicated workspaces. These come in four different flavours: isolated docker container, streaming desktop, shared virtual machine and dedicated virtual machine. The required docker container and virtual machine images are fetched from the private repository. This repository only contains docker images, and virtual machine images specifically created for the DTaaS platform.

The user workspaces are managed using compute infrastructure configuration tools \cite{beetz2021gitops}. These tools are being used to start and stop docker containers and virtual machines. All the user workspaces are started using provisioning scripts at the time of installation. An execution manager microservice will provide an application programming interface for on-demand provisioning of user workspaces. This microservice is still under development.

All the user workspaces have a unified interface via the web browser. In addition, these workspaces have Internet access. Thus a PT to DT bidirectional communication link is as simple as spawning a required client-server communication protocol software. It is possible to restrict the DT-PT and DT-Internet communication. A suitable network firewall configuration can easily enforce the necessary restrictions.

Users can also permit remote access to live DTs. There is already shared access to visualisation dashboards. With these two provisions, users can treat live DTs as service components in their own software systems.

\section{Future Work}\label{sec:future}

Quite a few development tasks are in-progress to enhance the functionality provided on the platform. We foresee three major areas of work.

First is the DT configuration and implementation. This is a significant research and implementation challenge, mostly because the domain is fuzzy and expanding so fast, that applying DSLs would be a moving target. There is a need to define a new DSL, or at least repurpose existing DSL to enforce the semantic restrictions on DT configuration. A more complex problem of configuration optimization awaits the researchers intending to scale up DT deployments in a cost-efficient manner. 

Second, we would like the DT platform tool to follow the DT lifecycle. 
In the beginning, the DT features provided are more or less standard, and focus is on those services that are application agnostic. The user provides DT assets (models, data, tools, functions), a configuration file, and gets a cloud based DT. The current platform is in this category.
Then, as the user of the DT learns more about the PT (through visualisation and basic historical simulation support), (s)he can develop basic state estimation algorithms that use simulation on historical data, such as the one introduced in \cite{Legaard2020}.
Then we can introduce more advanced state estimation algorithm such as the Kalman filtering, but these require slightly different models provided by the user. This introduces the challenge of keeping models consistent, since the models used for state estimation are similar to the models used for historical simulation, but have extra inputs. So we need a way to relate models in the platform, and keep them consistent.

 Third, is the on-demand management of user workspaces which frees up the execution resources and accommodates more users on the platform. A natural next step in this direction is the separation of DT lifecycle management from the user workspace. One user can create a DT, hand over the same to another user who will manage the other lifecycle phases of the given DT.
\section{Concluding Remarks}\label{sec:conclude}
There is a strong interest in the DT community to provide DTaaS to spread the user base of DTs. A typical DT lifecycle involves create, execute, save, analyse, evolve and terminate phases. Only software platforms developed with awareness of DT lifecycle can aspire to fulfil the DTaaS vision. Reusability of DT assets, creation of meaningful DT configuration, scalable deployment are key challenges in the development of DTaaS platforms.

In this paper, we have described some nuances in DT configuration that is valid in the context of reusable assets, shared infrastructure, and desired integration with external world. We propose one architecture for DTaaS software platform. With the aid of our implementation, we bring forth the potential benefits of our proposed DTaaS software architecture.

\subsection*{Source Code}
The source code for the project is available at:  \textit{https://github.com/INTO-CPS-Association/DTaaS}.

\section*{Acknowledgment}

This work has been partially supported by the EU Horizon 2020 projects DIGITbrain and HUBCAP  and the Poul Due Jensen foundation, 
as well as the RCN grants PeTWIN (294600) and SIRIUS (237898).

We thank in no particular order the successful discussions and feedback provided by the DIGITbrain technical coordination committee, Daniel Lehner, Mirgita Frasheri, Henrik Ejersbo, Gianmaria Bullegas and Omar Nachawati.

\IEEEtriggeratref{8}

\bibliographystyle{IEEEtran}
\bibliography{dtaas.bib}

\end{NoHyper}
\end{document}